\documentclass[10pt]{article}
\newcommand{\be}{\begin{equation}}
\newcommand{\ee}{\end{equation}}

\newcommand{\ba}[1]{\left(\begin{array}{#1}}
\newcommand{\ea}{\end{array}\right)}

\begin{document}
\baselineskip 14pt

\begin{center}
{\large\bf Spin squeezing and quantum correlations\\} 
\vspace{1em} 

A. R. Usha Devi$^{1,2}$ and Sudha$^{3}$    \\ 
{\small\it $^{1}$ Department of Physics, Bangalore University, Bangalore-560056, India.\\ 
$^{2}$ Inspire Institute Inc., Alexandria, Virginia, 22303, USA. \\
Email:arutth@rediffmail.com \\
 $^{3}$ Department of Physics, Kuvempu University, Shankaraghatta, Shimoga-577 451, India. \\ 
Email:arss@rediffmail.com \\}
\end{center}
\vspace{-1em}
\rule[-0.1em]{15.25cm}{0.5pt}
{\small
Recent years have witnessed revolutionary improvement in the production, manipulation, characterization and quantification of multiatom (multiqubit) states --  because of their promising applications in high precision atomic clocks, atomic interferrometry, quantum metrology and quantum information protocols. In the characterization and quantification of non-classical atomic correlations, two concepts namely, {\em spin squeezing} and {\em quantum entanglement} emerge. While spin squeezing originates from the uncertainty relation between atomic collective angular momentum operators, entanglement is a peculiar phenomena arising due to  superpositions in the multiparticle Hilbert space. Here, we outline different quantitative measures of spin squeezing and discuss the intrinsic relation of spin squeezing with quantum entanglement in multiqubit systems. } \\    
\rule[1em]{15.25cm}{0.5pt}

\section*{\normalsize 1 Introduction}

In conventional Ramsey spectroscopy, uncorrelated atoms are employed and the corresponding quantum noise (shot-noise)~\cite{Bachor} scales as $1/\sqrt{N}$, where $N$ is the number of atoms employed. An improved situation -- where noise scales as $1/N$, called the Heisenberg resolution limit  -- is envisaged~\cite{wineland, wineland2}  when quantum correlated atoms are used. Also, in atomic and optical interferometers -- which are standard tools for high precision phase resolution --  sensitivity can be improved (from the shot-noise limit $1/\sqrt{N}$ to Heisenberg limited resolution $1/N$) by employing entangled atoms and photons~\cite{Caves, yurke}. The advent of the rapidly progressing quantum information science (QIS) has brought forth the importance of entangled atoms/photons in several impressive applications. Understanding the nature of quantum correlations and quantifying them has thus captured the scenario during recent years. Well before the focus on quantifying entanglement begun, the notion of {\em spin squeezing} had caught much attention both in the context of low-noise spectroscopy as well as in high precision interferometry. While quantum correlation (entanglement) has its root in superposition  principle in multiparticle Hilbert space, squeezing originates from uncertainty principle. It may be pointed out here that squeezed radiation states have already received considerable attention as basic ingredients of non-classicality  in the theory of continuous variable QIS~\cite{cv}. On the other hand, in discrete variable QIS, which is based mainly on two-level systems (atoms with two internal levels or qubits), there have been intense efforts recently~\cite{sosn, ushamallesh, usha, sanders1, sanders2, ushasudha, akrusha,gtoth} to understand the intrinsic link between spin squeezing and entanglement. In this paper, we report theoretical efforts on developing quantitative measures of spin squeezing and also relate spin squeezing with  quantum entanglement in multiatomic (multiqubit) systems.

The article is organized as under. In Section~2, we give a brief review of the concept and origins of the definitions of spin squeezing proposed by Kitagawa-Ueda~\cite{ku}, Wineland~\cite{wineland2}. 
Section~3 illustrates, with the help of two simple  examples, the need for a local invariant spin squeezing criteria and the inability of Kitagawa-Ueda spin squeezing criteria to capture quantum correlations -- except in the case of composite systems  obeying exchange symmetry.  Section~4 deals with the development of the locally invariant spin squeezing and also an operational approach to evaluate the corresponding spin squeezing parameters using the state parameters of a multiqubit system. The connection between spin squeezing and quantum entanglement is discussed in Section~5. In particular, the relationship between the spin squeezing parameter and concurrence for pure two-qubit states is outlined here. The importance of local invariant spin squeezing criteria to distinguish between separable and entangled multiqubit states is also discussed. We also analyze a  relationship between one of the entanglement invariants of a symmetric two qubit state with the collective spin-squeezing parameter associated with an intrinsically symmetric multiqubit state. Section~6 contains concluding remarks.

\section*{\normalsize 2 Spin squeezing}

The notion of squeezing which involves reduction in the variance (uncertainty) of an observable
below a standard quantum limit has been studied \cite{kimble, stoler, NM} in the literature for 
bosonic fields. The squeezed states of radiation have received due attention in
the last decade and investigations so far on bosonic squeezed states  have focused on various aspects such as the nonclassical features associated with them, on possible ways and means of generating
them and also on the practical applications of these states to achieve minimum noise in amplifiers
and in optical interferometers~\cite{kimble, stoler, NM}. The concept of squeezing was later extended to non-canonical systems such as spin \cite{wineland, ku}.  
Although squeezing is unambiguously defined in the case of bosonic systems \cite{kimble, stoler, NM}, its definition in the context of spin required 
careful consideration. There have been several definitions of spin squeezing~\cite{wineland, wineland2, sosn, ku, wkz, gsapuri, puri} which depend on the context in which they are defined. We outline below the original concept of spin squeezing proposed by Kitagawa-Ueda~\cite{ku}, Wineland~\cite{wineland2}. 

The components of the spin operator $\vec{J}$ satisfy the commutation relations (here we have employed $\hbar=1$)
\begin{equation}
\label{com1}
\left[ J_x,\,J_y  \right]=iJ_z;\, x,\,y,\,z\  \mbox{cyclic}  
\end{equation}
and hence obey the uncertainty relationships 
\begin{equation}
\label{mu1}
\Delta J^2_x \Delta J^2_y \geq  \frac{\langle J_z\rangle^2}{4} ; \   x,\,y,\,z\  \mbox{cyclic}. 
\end{equation}
(Here, $\Delta A^2=\langle A^2\rangle-\langle A\rangle^2$). 
 
Comparison with the quadrature squeezing criterion for the radiation fields viz., $\Delta x^2<\frac{1}{2}$ or $\Delta p_x^2<\frac{1}{2}$ --  while maintaining  the uncertainty product $\Delta x^2\Delta p_x^2\geq \frac{1}{4}$ --   naturally suggests that a spin state could be regarded as squeezed if $\Delta J^2_x$ or $\Delta J^2_y$ is smaller than $\frac{\vert \langle J_z \rangle \vert}{2}.$
Indeed, this has been employed earlier as the spin squeezing criterion in the literature \cite{wkz}. However, this criterion was first critically examined by Kitagawa and Ueda \cite{ku}, who pointed out that such a definition is coordinate dependent in the sense that a spin state which is not squeezed in a given coordinate frame of reference will  be squeezed in a rotated coordinate frame of reference. In an attempt to arrive at a {\em physical} criterion of spin squeezing, which is frame independent, Kitagawa and Ueda \cite{ku} considered the model in which a spin-$j$ state is thought of as being built out of $N=2j$ elementary spin-$\frac{1}{2}$ systems. A coherent spin state (CSS)~\cite{Arechchi} 
\begin{eqnarray} 
  \label{scs}
   \vert \theta,\phi\rangle&=&e^{\tau\, J_+-\tau^*J_-}\, 
   \vert j=N/2, -j=-N/2\rangle \nonumber \\ 
   &=& \sum_{k=0}^{N}\, \sqrt{^{N} C_{k}}\, \left(\cos\frac{\theta}{2}\right)^{N-k}\, \left(\sin\frac{\theta}{2}\right)^{k} \  e^{ik\phi}\  
   \left\vert j=\frac{N}{2}, k-\frac{N}{2}\right\rangle, 
  \end{eqnarray}
  (where $\tau=\frac{\theta}{2}\, e^{i\phi}$, 
$0\leq \phi\leq 2\pi$, $0\leq \theta\leq \pi$;  $J_\pm=\sum_{i=1}^{N}\, \sigma_{i\pm}$ are the collective spin ladder operators)  
is then recognized as a state in which all the $N$ elementary spinors point in the same direction $\hat n_0(\theta,\, \phi)$. Apart from being an eigenstate of the spin component $\vec{J}\cdot \hat n_0$ with eigenvalue $j=-N/2$, the coherent spin state satisfies 
the minimum uncertainty relationship namely Eq.~(\ref{mu1}) with equality sign; the uncertainties on the LHS of Eq.~(\ref{mu1}) -- found to be  $\frac{N}{4}$ -- are equally distributed over any two orthogonal spin components normal to the direction of spin $\hat n_0$. As this state consists of $N$ internal spin-$\frac{1}{2}$ systems,  all pointing in the same direction $\hat n_0$, Kitagawa and Ueda~\cite{ku} point out that there are no quantum correlations in a CSS. However,  CSS is shown to exhibit spin squeezing in some rotated frames of reference~\cite{ku}, if one confines to the definition of spin squeezing based merely on the uncertaintly relation Eq.~(\ref{mu1}).  
In order to define a rotation invariant spin squeezing criterion,  Kitagawa and Ueda~\cite{ku} first identified a mean spin direction 
\begin{equation}
{\hat n_0}=\frac{\langle  \vec J  \rangle}{\vert \langle  \vec J  \rangle \vert}, \ \ \mbox{where} \ \ \vert \langle  \vec J  \rangle \vert=\sqrt{\langle  \vec J  \rangle \cdot \langle  \vec J  \rangle}
\end{equation}
where the collective spin operator $\vec J$ for an $N$-qubit system is defined by ${\vec  J}=\frac{1}{2} \sum_{i=1}^N  \vec \sigma_i$ with 
${\vec \sigma_i}=(\sigma_x,\,\sigma_y,\,\sigma_z)$  being the Pauli operator of the $i$th qubit. Associating a mutually orthogonal basis set 
$({\hat n}_{\bot}, {\hat n}'_{\bot},{\hat n_0})$ the collective operators given by 
\begin{equation}
J_\mu={\vec J}\cdot {\hat n}_\mu, \  \mbox{where} \ \ {\hat n}_\mu \equiv({\hat n}_{\bot}, {\hat n}'_{\bot},{\hat n_0}), 
\end{equation}
can be easily seen to satisfy the angular momentum commutative relations. Kitagawa and Ueda~\cite{ku} proposed that a multiqubit state be regarded as spin squeezed if the minimum of $\Delta J_\bot$ (or equivalently $\Delta J'_\bot$) of a spin component normal to the 
mean spin direction is smaller than the standard quantum limit $\frac{\sqrt N}{2}$  of the CSS.  In other words, 
\[
{\rm spin\ squeezing}\ \Rightarrow (\Delta J_\bot)_{\mbox{min}} \leq \frac{\sqrt N}{2}\Rightarrow \frac{2(\Delta J_\bot)_{\mbox{min}}}{\sqrt{N}}\leq 1, 
\]
and hence  spin squeezing parameter -- a quantitative measure of spin squeezing -- incorporating this feature may be defined as, 
\begin{equation}
\label{ku}
\xi_1=\frac{2(\Delta J_\bot)_{\mbox{min}}}{\sqrt{N}}.
\end{equation}
 A spin squeezed state obviously has the corresponding parameter $\xi_1<1$. 

In the context of Ramsey spectroscopy with a sample of $N$ two-level atoms, Wineland \textit{et.al}~\cite{wineland2} showed that the frequency resolution depends on the parameter
\begin{equation}
\label{wineland}
\xi_2=\frac{\sqrt{N}(\Delta J_\bot)_{\mbox{min}}}{\vert\langle J_0 \rangle\vert}=\frac{N\xi_1}{2\vert\langle J_0 \rangle\vert}
\end{equation}
and the spin squeezing manifested by $\xi_2<1$ leads to reduction in the frequency noise. 

As $\vert\langle J_0 \rangle\vert_{\rm max}=\frac{N}{2}$ holds for CSS, it is clear that $\xi_1=\xi_2$ for CSS. Also, as $\vert{\langle J_0 \rangle}\vert<\frac{N}{2}$ for $N$-qubit states other than coherent ones, $\xi_2<1\Rightarrow \xi_1<1$. In other words, $\xi_1<1$ is a necessary, but not a sufficient criterion for $\xi_2<1$ i.e., not all states that are spin-squeezed in the Kitagawa-Ueda sense are to be spin-squeezed in the Wineland sense. All these together imply that the condition for squeezing put forth by 
Wineland \textit{et.al} \cite{wineland2} is a stricter condition -- when compared with that proposed by Kitagawa and Ueda \cite{ku}. 

\section*{\normalsize 3 Exchange symmetry, local Invariance and spin squeezing} 

Spin squeezing, in the original sense, is defined for multiqubit states that are invariant under the exchange of particles~\cite{ulamku}. Such states, the so-called \textit{symmetric states } belong to the maximal multiplicity subspace of the collective angular momentum operator $\vec J$.  The possibility of extending the concept of spin squeezing to multi-qubit systems that are not necessarily symmetric under interchange of particles and that are accessible not just to collective operations but also to local operations was explored in Ref.~\cite{usha}. This requires a criterion for spin squeezing that exhibits invariance under local unitary operations on the qubits. At this juncture, it is important to notice that the spin squeezing parameters given by Eqs.~(\ref{ku}), (\ref{wineland}) are not invariant under arbitrary local unitary transformations on the qubits. In order to see this, let us consider the following example of two qubits:

\begin{enumerate}
\item Consider the state \cite{usha}
\begin{equation}
\vert \psi \rangle= \cos\theta\, \vert 0_1,1_2\rangle+\sin\theta\, \vert 1_1,0_2 \rangle  
\end{equation}
for which $\langle {\vec J} \rangle=0$ and therefore it is not possible to  define spin squeezing for this system. However, under a local unitary operation $U_1\otimes U_2=I\otimes\sigma_x$ (where $I$ denotes $2\times 2$ identity matrix and $\sigma_x$, the Pauli spin operator), the state $\vert \psi\rangle$ gets transformed to 
\begin{equation}
\vert \psi' \rangle= \cos\theta\, \vert 0_1,0_2\rangle+\sin\theta\, \vert 1_1,1_2 \rangle.  
\end{equation}
It may be readily verified that the average spin in this state is given by $\langle\vec{J}\rangle=(0, 0, \cos 2\theta)$; the variances   $(\Delta J^2_\bot)_{\mbox{min}}=\frac{1-\vert\sin 2\theta\vert}{2}$.  So, one obtains the  result that $\vert \psi' \rangle$ is spin squeezed with the squeezing parameters given by 
\begin{equation}
\xi_1=\sqrt{1-\vert \sin 2\theta\vert}\leq 1,\ \ \xi_2=\frac{1}{\sqrt{1+\vert \sin 2\theta\vert}}\leq 1. 
\end{equation}
In other words, the spin squeezing defined by either criterion gets modified by a local unitary transformation -- which corresponds to a basis change in the individual qubit spaces.  

Having shown that $\xi_1$, $\xi_2$ are not local unitary invariants, we will show, through the following example that spin squeezing criterion given by Kitagawa and Ueda \cite{ku} works well only for states exhibiting exchange symmetry. 
\item Consider the following non-symmetric, uncorrelated state of two qubits:
\begin{equation}
\psi=\left[\frac{\sqrt{3}}{2}\vert 0_1\rangle + \frac{1}{2} \vert 1_1\rangle\right] \otimes  \left[\frac{\sqrt{3}}{2} \vert 0_2\rangle -\frac{1}{2}\vert 1_2\rangle\right].
\end{equation}
The mean spin lies along the z-axis with $\vert\langle J_0 \rangle\vert=\frac{1}{2}$ and $(\Delta J^2_\bot)_{\mbox{min}}=\Delta J^2_x=\frac{1}{8}$. Hence one gets $\xi_1=\frac{1}{2}$ for the above state implying that the state exhibits spin-squeezing. In general, non-symmetric product states of the form 
$\vert \psi_1\rangle \otimes \vert \psi_2 \rangle$ can also exhibit spin squeezing -- which is undesirable because a product state obviously does not possess any quantum correlations.  Thus, it can be concluded that when applied to  non-symmetric  states,  the Kitagawa-Ueda spin squeezing criteria leads to  unphysical results.  The symmetric product state  
\begin{equation}
\psi=\left[\frac{\sqrt{3}}{2}\vert0_1\rangle+\frac{1}{2}\vert 1_1\rangle \right]\otimes \left[\frac{\sqrt{3}}{2}\vert0_2\rangle+\frac{1}{2}\vert 1_2\rangle \right]
\end{equation}
(being a two qubit spin coherent state), leads to $\xi_1=1=\xi_2$ which clearly imply that the symmetric product state is {\em not} spin-squeezed.
\end{enumerate}

The above two simple examples illustrate the need for identifying a {\em local invariant} criterion of spin squeezing and also point towards the lack of a clear understanding of  spin squeezing in  non-symmetric states.  A local invariant version of the spin squeezing criteria,  which yield satisfactory results when applied  to multiqubit states that do not necessarily obey exchange symmetry, has been developed in Ref.~\cite{usha}.   The development and significance of the local invariant spin-squeezing criteria  is detailed in the next section. 

\section*{\normalsize 4 Local invariant criteria for spin squeezing in multiqubit systems}    

As a starting point for developing local invariant criteria for spin squeezing, Usha Devi \textit{et.al} \cite{usha} considered a set of mutually orthogonal unit vectors $({\hat n}_{i\bot},{\hat n}'_{i\bot}, {\hat n}_{i0})$ associated with $i$th qubit of the $N$-qubit system, $i=1,\,2,\,\ldots N$. Here ${\hat n}_{i0}$ is a unit vector along mean spin direction of the $i$th qubit given by  
\be
{\hat n}_{i0}=\frac{{\vec\sigma}_i}{\vert {\langle\vec\sigma}_i\rangle \vert}, \ \ \vert {\langle\vec\sigma}_i \rangle\vert=\sqrt{\langle{\vec\sigma}_i\rangle\cdot \langle{\vec\sigma}_i\rangle}
\ee
and ${\hat n}_{i\bot}$, ${\hat n}'_{i\bot}$ are mutually orthogonal unit vectors  perpendicular to the  mean spin direction ${\hat n}_{i0}$.  Three collective {\em angular momentum} operators defined through, 
\be
\label{gen}
{\cal J}_\bot=\frac{1}{2}\sum_{i=1}^N \vec{\sigma}_i \cdot \hat{n}_{i\bot}, \ \ {\cal J}'_\bot=\frac{1}{2}\sum_{i=1}^N \vec{\sigma}_i \cdot \hat{n}'_{i\bot},\ \ {\cal J}_0=\frac{1}{2}\sum_{i=1}^N \vec{\sigma}_i \cdot \hat{n}_{i0} 
\ee 
satisfy commutation relations  $$[{\cal J}_\bot, {\cal J}'_\bot]=i{\cal J}_0$$ 
and hence they obey the uncertainty relations
\be
\label{unc}
(\Delta{\cal J}_\bot)({\Delta{\cal J}'_\bot})\geq \frac{1}{2}\, \vert\langle {\cal J}_0 \rangle\vert.
\ee 
Spin squeezing of $N$ qubits, resulting from the possibility of
redistributing the fluctuations unevenly between ${\cal J}_\bot$ and ${\cal J}'_\bot$, without
violating the Heisenberg uncertainty relation Eq.~(\ref{unc}), is
formulated through a quantity which exploits the inequality
$(\Delta {\cal J}_\bot)_{\mbox{min}} < \sqrt{\vert\langle {\cal J}_0 \rangle \vert/2}$. 
Here the subscript ``min'' denotes the minimum value of $\Delta {\cal J}_{\bot}$ (or $\Delta {\cal J}'_\bot$)
achieved by appropriate choice of directions ${\hat n}_{i\bot}$ (or  ${\hat n}'_{i\bot}$). 

Local invariant spin squeezing parameters $\tilde{\xi}_1$, $\tilde{\xi}_2$ --  analogous to those proposed by Kitegawa-Ueda and Wineland -- may then be defined for an arbitrary multiqubit system  as follows:  
\be
\label{kuw2}
\tilde{\xi}_1=\frac{2 {(\Delta {\cal J}_\bot)}_{\mbox {min}}}{\sqrt N}, \ \ \tilde{\xi}_2=\frac{\sqrt{N}{(\Delta {\cal J}_\bot)}_{\mbox {min}}}{\vert \langle {\cal J}_0\rangle \vert}
\ee 
It may be readily verified that $N$-qubit CSS and the set of all $N$ qubit states related to CSS by local unitary operations, satisfy the minimum uncertainty relationship $(\Delta{\cal J}_\bot)({\Delta{\cal J}'_\bot})=\frac{1}{2}\langle {\cal J}_0 \rangle$ with $(\Delta {\cal J}_\bot)=(\Delta {\cal J}'_\bot)=\frac{\sqrt N}{2}$ and $\vert \langle {\cal J}_0\rangle\vert=\frac {N}{2}$. The definitions 
of spin squeezing imply that any arbitrary $N$-qubit state with $\tilde\xi_k<1$ do exhibit reduced
fluctuations for the generalized collective angular momentum operators $\cal J_\bot$, 
below the standard quantum limit $\sqrt{N}/2$. 

The local invariant nature of the spin squeezing parameters $\tilde{\xi}_k$ may be  established explicitly as follows: 

On expressing $(\Delta {\cal J}_\bot)^2_{\mbox {min}}$ as (see Eq.~(\ref{gen})) 
\be
\label{varJ}
(\Delta {\cal J}_\bot)^2_{\mbox {min}}=\frac{1}{4}\left[ N+2\left(\sum_{i=1}^N \sum_{j>i=1}^N { {\hat n}_{i\bot}}^T{\cal T}^{(ij)}{\hat n}_{j\bot}  \right)_{\mbox {min}} \right]
\ee
where  
\be
{\cal T}^{(ij)}_{\alpha\beta}=\langle \sigma_{i\alpha} \sigma_{j\beta} \rangle, \ \ \alpha,\, \beta=x,\,y,\,z 
\ee
are the matrix elements of the $3\times 3$ correlation matrix  ${\cal T}^{(ij)}$ assoicated with the pair of qubits labeled $i,j$. 
Under local unitary transformations $U_1\otimes U_2\otimes \ldots \otimes U_N$,  the unit vectors ${\hat n}_{i\bot}$ and the correlation matrix ${\cal T}^{(ij)}$ transform as \cite{rh}
\be
{\hat n}'_{i\bot}=O_i {\hat n}_{i\bot}, \ \ \ {\cal T'}^{(ij)}=O_i {\cal T}^{(ij)}{O_j}^T
\ee
and hence the quantity ${ {\hat n}^T_{i\bot}}{\cal T}^{(ij)}{\hat n}_{i\bot}$ remains unchanged under local unitary transformations. (Here, we have used the well-known  fact that for every $2\times2$ unitary transformation $U_i$ on a $i$th qubit, there corresponds a unique
$3\times 3$ real orthogonal rotation matrix $O_i$).
The expectation value $\langle {\cal J}_0 \rangle=\frac{1}{2} \sum_{i=1}^N \vert {\langle \vec \sigma}_{i0}\rangle \vert$ is a local invariant as is evident from the fact that the quantities $\vert \langle{\vec \sigma}_{i0}\rangle \vert$, being the magnitudes of the spins of individual qubits, remain the same under local rotations. As the expressions for $\tilde{\xi_1}$, $\tilde{\xi_2}$ involve only the quantities  $(\Delta J_\bot)_{\mbox {min}}$ and $\langle {\cal J}_0 \rangle$, their local invariance readily establishes the local unitary invariance of $\tilde {\xi_1}$, $\tilde {\xi_2}.$ 

Though the minimization of $(\Delta {\cal J}_\bot)$ over all the directions ${\hat n}_{i\bot}$
appears to be non-trivial, local invariance of ${\tilde \xi}_k$ leads to a simplified mathematical analysis for multi-qubit
systems.  An operational approach for evaluating $\tilde \xi_k$ making use of the local invariant property of the squeezing parameters is outlined in the next subsection.  

\subsection*{\normalsize 4.1 Operational approach towards the evaluation of the spin-squeezing parameters $\tilde \xi_k$} 

The evaluation of the parameters $\tilde\xi_k, k=1,2$, it is essential that the minimum value of the variance ${(\Delta {\cal J}_\bot)}_{\mbox{min}}$ be computed. In order to take up this task, it is convenient to group the multiqubit states into two categories;

\begin{itemize}
\item[(i)]  States that are either symmetric under interchange of particles or those that are convertible to states obeying exchange symmetry through local unitary transformations. 
\item[(ii)] States that are {\em not} intrinsically symmetric under exchange of particles. 
\end{itemize}
For instance, if we consider a product state with each qubit
being in a different state, one can always find a unitary
operator $U_1\otimes U_2\otimes \ldots \otimes U_N$ to transform this state into a symmetric product state  
$\vert 0 \rangle \otimes \vert 0 \rangle\otimes \ldots \vert 0\rangle$. In other words, the set of all product qubit states are local unitarily equivalent to spin coherent states and hence fall into the category (i). There are innumerably many multiqubit states that are intrinsically non-symmetric and hence are not convertible to symmetric multiqubit states under any set of local unitary operations. Such states come under category (ii). 

When we consider the states belonging to category (i), it is easy to see that one can appeal to a common orientation  
\be
\label{co} 
({\hat n}_{i\bot}, {\hat n}'_{i\bot}, {\hat n}_{i0})\longrightarrow ({\hat n}_{\bot}, {\hat n}'_{\bot}, {\hat n}_{0})
\ee
with the help of local unitary operations $U_1\otimes U_2\otimes \ldots \otimes U_N$, such that the  bipartite reduced density matrices of  all the $\frac {N(N-1)}{2}$  pairs of qubits, extracted from an intrinsically symmetric state, are all identical, with magnitude of individual qubit orientation vectors 
$\vert\langle \vec{\sigma}_i\rangle\vert\equiv s_0$ and the two qubit correlation matrices ${\cal T}^{(ij)}\equiv {\cal T}$ being identically same  for all the constituent qubits. Therefore, for all the  $\frac {N(N-1)}{2}$  pairs of qubits, extracted from an intrinsically symmetric $N$ qubit state, we have \be
\label{tmin}
\sum_{i=1}^N \sum_{j>i=1}^N { {\hat n}^T_{i\bot}}{\cal T}^{(ij)}{\hat n}_{i\bot}   \longrightarrow \frac{N(N-1)}{2}{ {\hat n}^T_{\bot}}{\cal T}{\hat n}_{\bot}. 
\ee 
under a suitable local unitary operation on the qubits. We also obtain the mean value  $\langle {\cal J}_0 \rangle$ of the collective spin operator
as, 
\be
\langle {\cal J}_0 \rangle=\sum_{i=1}^{N}\vert \langle \vec{\sigma}_i \rangle\vert \equiv \frac{N\, s_0}{2}.
\ee
The local invariant spin squeezing parameters $\tilde \xi_k$ of Eq.~(\ref{kuw2}) reduce to the following
a simple form: 
\begin{eqnarray}
\label{xitil1}
{\tilde \xi_1}&=&\left[1+(N-1)({ {\hat n}^T_{\bot}}{\cal T}{\hat n}_{\bot})_{\mbox{min}}\right]^{1/2}  \\
\label{xitil2}
{\tilde \xi_2}&=&\frac{\left[1+(N-1)({ {\hat n}^T_{\bot}}{\cal T}{\hat n}_{\bot})_{\mbox{min}}\right]^{1/2}}{s_0}=\frac{N\,{\tilde \xi_1}}{2\langle{\cal J}_0\rangle}. 
\end{eqnarray}
Specifying the qubit orientation direction $\hat n_0$  to be
the $z$-axis, without any loss of generality, such that ${\hat n}_{\bot}=\left( \cos\theta, \, \sin\theta,\,0 \right)$, the quadratic form $({ {\hat n}^T_{\bot}}{\cal T}{\hat n}_{\bot})_{\mbox{min}}$ is readily simplified as, 
\begin{eqnarray}
\label{nmin}
({ {\hat n}^T_{\bot}}{\cal T}{\hat n}_{\bot})_{\mbox{min}}&=&{\mbox{min}}_{\theta}\left( {\cal T}_{xx}\,\cos^2\theta+{\cal T}_{yy}\,\sin^2\theta+ {\cal T}_{xy}\sin 2\theta \right)\nonumber \\
&=&\frac{1}{2}\left[ ({\cal T}_{xx}+{\cal T}_{yy})-\sqrt{({\cal T}_{xx}-{\cal T}_{yy})^2+4{\cal T}^2_{xy}} \,\right]
\end{eqnarray}
and hence the locally invariant spin squeezing parameters $\tilde \xi_k$ are obtained, in an operational form, on substituting Eq.~(\ref{nmin}) into  Eqs.~(\ref{xitil1}), (\ref{xitil2}). In other words, the locally invariant spin squeezing parameters $\tilde \xi_k$ may be evaluated in terms of   the elements of the correlation matrix ${\cal T}$ of any random pair of qubits, and the single qubit parameter $\vert\langle \vec{\sigma}_i\rangle\vert=s_0$. 

The local invariant spin squeezing parameters $\tilde \xi_k$ reduce to
the spin squeezing parameters $\xi_k$ of Eqs.~(\ref{ku}), (\ref{wineland}) for symmetric multi-
qubit systems. In other words, the parameters $\tilde \xi_k$ provide a natural
characterization of collective spin squeezing in intrinsically symmetric multi-qubit
systems -- equipped with an additional feature that local unitary operations
on the qubits do not increase (reduce) them.

For intrinsically non-symmetric multiqubit states, any common qubit orientation Eq.~(\ref{co}) 
 -- achievable with the help of local unitary operations on   the qubits -- does not lead to identical
bipartite density matrices for the qubit pairs. However, one can obtain an
analogous approach to evaluate the spin squeezing parameters in a similar fashion. By appealing to an appropriate local unitary transformation to orient all the qubits along a common frame of reference as in Eq.~(\ref{co}), one obtains,  
\be
\sum_{i=1}^N \sum_{j>i=1}^N { {\hat n}^T_{i\bot}}{\cal T}^{(ij)}{\hat n}_{j\bot}   \longrightarrow { {\hat n}^T_{\bot}} \texttt{T}{\hat n}_{\bot},
\ee
where $\texttt{T}=\sum_{i=1}^N \sum_{j>i=1}^N {\cal T}^{(ij)}$ is the sum of transformed correlation matrices of all
the $\frac{N(N-1)}{2}$ pairs of qubits. The quadratic form ${ {\hat n}^T_{\bot}} \texttt{T}{\hat n}$ 
can be written in the symmetric manner as, 
\be
{ {\hat n}^T_{\bot}} \texttt{T}{\hat n}_{\bot}={ {\hat n}^T_{\bot}}\frac{ (\texttt{T}+ \texttt{T}^T)}{2} {\hat n}_{\bot} = { {\hat n}^T_{\bot}} \texttt{S}{\hat n}_{\bot};\ \ \frac{\texttt{T}+ \texttt{T}^T}{2}={\texttt S}. 
\ee
Thus, the parameters $\tilde \xi_k$ (see  Eqs.~(\ref{kuw2}), (\ref{varJ}))  can be expressed as 
\begin{eqnarray}
\label{cat2}
{\tilde \xi_1}&=& \frac{1}{2\sqrt{N}}\left[ N+ \left({ {\hat n}^T_{\bot}} \texttt{S}{\hat n}_{\bot}\right)_{\mbox{min}} \right]^{1/2} \nonumber \\ 
{\tilde \xi_2}&=& \frac{\sqrt{N}}{2\langle{\cal J}_0 \rangle}\left[ N+ \left({ {\hat n}^T_{\bot}} \texttt{S}{\hat n}_{\bot}\right)_{\mbox{min}} \right]^{1/2}
\end{eqnarray}
The minimum value of the quadratic form ${ {\hat n}^T_{\bot}} \texttt{S}{\hat n}_{\bot}$ is readily identified to be,  
\be
\left({ {\hat n}^T_{\bot}} \texttt{S}{\hat n}_{\bot}\right)_{\mbox{min}}=
\frac{1}{2}\left[ (\texttt {S}_{xx}+\texttt {S}_{yy})-\sqrt{ \left( \texttt {S}_{xx}-\texttt {S}_{yy}\right)^2+ 4 \texttt {S}^2_{xy}}\,\right]
\ee
Finally, on substituting for $\left({ {\hat n}^T_{\bot}} \texttt{S}{\hat n}_{\bot}\right)_{\mbox{min}}$ from the above equation into Eq.~(\ref {cat2}),  the locally invariant spin squeezing parameters $\tilde \xi_k$ for multiqubit states that are not intrinsically symmetric, are cast into a computable form.  

The generalized collective spin operators (defined through Eq.~(\ref{gen})) of a
multi-qubit spin squeezed state can be represented geometrically by an
elliptical cone of local invariant height $\langle{\cal J}_0 \rangle$ centered about the z-axis
(common qubit orientation direction in the case of symmetric multiqubit states), semi-minor and semi-major axes of
the ellipse being $(\Delta {\cal J}_\bot)_{\rm min}$ and the corresponding value $\Delta {\cal J}'_\bot$ respectively. In
contrast, the coherent spin state is depicted by a circular cone. This
geometric picture of the spin squeezed state versus that of a spin coherent state illustrates collective spin-squeezing feature 
in non-symmetric states.

We now proceed further to discuss the relationship between spin squeezing and entanglement.

\section*{\normalsize 5 Relationship between spin squeezing and quantum entanglement}
A deeper understanding  between spin squeezing and quantum entanglement has been explored in recent literature~\cite{sosn,usha,sanders1,sanders2, ulamku} and it has been established  that the presence of spin squeezing essentially reflects pairwise entanglement in multiqubit systems.  Usha Devi {\textit et.al} \cite{usha} illustrated that local invariant  spin squeezing criteria necessarily imply quantum entanglement. This section is devoted to explicitly demonstrate this feature.  

\subsection*{\normalsize 5.1 Spin squeezing and quantum entanglement in pure two qubit states}

An arbitrary two-qubit pure state given by 
\be
\vert \phi \rangle=\alpha\,\vert 0_1,0_2\rangle+\beta\,\vert 0_1,1_2\rangle+\gamma\, \vert 1_1,0_2\rangle+\delta \, \vert 1_1,1_2\rangle 
\ee   
is equivalent, up to local unitary equivalence, to
\begin{eqnarray}
\vert \phi' \rangle&=&\lambda_1\, \vert 0'_1,0'_2\rangle+\lambda_2\,\vert 1'_1,1'_2\rangle \nonumber \\
0\leq \lambda_1,\,\lambda_2\leq 1,& & \lambda_1^2+\lambda_2^2=1.   
\end{eqnarray}
Here $\lambda_1$, $\lambda_2$ are the Schmidt coefficients ($\lambda^2_1$ and $\lambda^2_2$ are the eigenvalues of the reduced single qubit density matrices) and they are related to $\alpha,\,\beta,\, \gamma,\, \delta$ through 
\[ 
\lambda_1^2=\frac{1}{2}\left[ 1+\sqrt{1-4\vert (\beta\gamma-\alpha\delta)\vert^2}\,\right],\ \  \lambda_2^2=\frac{1}{2}\left[ 1-\sqrt{1-4\vert (\beta\gamma-\alpha\delta)\vert^2}\,\right].
\] 
Identifying that  
\be
\langle {\cal J}_0 \rangle=\vert \lambda_1^2-\lambda_2^2 \vert,\ \ \texttt{T}=\mbox{diag}\,(2\lambda_1\lambda_2,\, -2\lambda_1\lambda_2,\,0)  
\ee
the local invariant spin squeezing parameters are found to be 
\be
\label{sq3}
\tilde \xi_1=\sqrt{1-2\lambda_1\lambda_2},\ \ \tilde \xi_2=\frac{\sqrt{1-2\lambda_1\lambda_2}}{\vert \lambda_1^2-\lambda_2^2 \vert}.
\ee
Here, we employ concurrence $0\leq {\cal C}\leq 1$ -- a well-known measure of two-qubit entanglement~\cite{wootters} -- to verify the relation between spin squeezing and entanglment. Using the fact that concurrence ${\cal C}$ of an arbitrary pure two qubit state is related to the Schmidt coefficients $\lambda_1$, $\lambda_2$ through 
${\cal C}=2\lambda_1\lambda_2=2\vert(\beta\gamma-\alpha\delta)\vert$, and expressing $\vert(\lambda_1^2-\lambda_2^2) \vert=\sqrt{1-{\cal C}^2}$ 
in Eq.~(\ref{sq3}), we obtain, 
\be
\tilde \xi_1=\sqrt{1-{\cal C}},\ \ \tilde \xi_2=\frac{1}{1+{\cal C}}.
\ee
These relations reveal that local invariant spin squeezing and quantum entanglement are equivalent for arbitrary two-qubit pure states. A more general argument to establish the connection between spin squeezing and entanglement  is outlined in the next subsection.  

\subsection*{\normalsize 5.2 Entanglement and spin-squeezing in multiqubit states} 

A multiqubit system is said to be entangled iff its density matrix cannot be expressed in a fully separable form  
\be
\label{sp}
\rho^{(\rm sep)}=\sum_k\,p_k\, \rho_1^{(k)}\otimes \rho_2^{(k)}\otimes \ldots \otimes \rho_N^{(k)},\ \ \sum_k\, p_k=1.  
\ee
where $\rho_i^{(k)}$ are density matrices of the $i$th qubit. 
We now proceed to  show that a spin squeezed $N$-qubit state -- characterized by $\tilde{\xi}_2<1$ -- is necessarily entangled~\cite{usha}.

In a separable state Eq.~(\ref{sp}), the variance of ${\cal J}_{\bot}$ (see Eq.~(\ref{varJ})) is evaluated as under; 
\begin{eqnarray}
\label{in1}
(\Delta {\cal J}_\bot)^2&=&\frac{N}{4}+\frac{1}{4}\sum_k\,p_k\,\sum_{i=1}^N \sum_{j\neq i=1}^N \langle \vec\sigma_i\cdot\hat n_{i\bot}\rangle_k \langle \vec\sigma_j\cdot\hat n_{j\bot}\rangle_k \nonumber \\
&=&\frac{N}{4}+\frac{1}{4}\sum_k\,p_k\,\sum_{i} \sum_{j} \langle \vec\sigma_i\cdot\hat n_{i\bot}\rangle_k \langle \vec\sigma_j\cdot\hat n_{j\bot}\rangle_k-\frac{1}{4}\sum_k\, p_k\,\sum_i \langle \vec\sigma_i\cdot\hat n_{i\bot}\rangle_k^2\nonumber \\
&=&\frac{N}{4}+\frac{1}{4}\sum_k p_k\,\left( \sum_i \langle \vec\sigma_i\cdot\hat n_{i\bot}\rangle_k \right)^2-\frac{1}{4} \sum_{k} p_k\,\sum_i\langle \vec\sigma_i\cdot\hat n_{i\bot}\rangle_k^2 \nonumber \\
&&\geq \frac{N}{4}-\frac{1}{4}\sum_k\, p_k\,\sum_i \langle \vec\sigma_i\cdot\hat n_{i\bot}\rangle_k^2 
\end{eqnarray}
Using the condition  
\be
\langle {\vec \sigma_i\cdot {\hat n}_{i\bot}} \rangle^2_k+\langle {\vec \sigma_i\cdot {\hat n}'_{i\bot}} \rangle^2_k+\langle {\vec \sigma_i\cdot {\hat n}_{i0}} \rangle^2_k \leq 1
\ee  
one obtains 
\begin{eqnarray}
\label{in2}
\frac{N}{4}-\frac{1}{4}\sum_k\, p_k\,\sum_i \langle \vec\sigma_i\cdot\hat n_{i\bot}\rangle_k^2&\geq& \frac{1}{4}\sum_k\, p_k\,\sum_i [\langle {\vec \sigma_i\cdot {\hat n}'_{i\bot}} \rangle^2_k+\langle {\vec \sigma_i\cdot {\hat n}_{i0}} \rangle^2_k]. 
\end{eqnarray} 
 Expressing, 
 \be
 \label{min}
 \frac{1}{4}\sum_k\, p_k\,\sum_i \langle {\vec \sigma_i\cdot {\hat n}_{i0}} \rangle^2_k=\frac{1}{N}\, \left\{\sum_{i=1}^{N}\sum_{k}\sqrt{p_k}^2\right\} 
 \left\{\sum_{i=1}^{N}\sum_{k}\left(\sqrt{p_k}\langle {\vec \sigma_i\cdot {\hat n}_{i0}} \rangle_k\right)^2\right\} 
 \ee
 and using the Schwarz inequality $\left(\sum_{l} A_l\, B_l\right)^2\leq \sum_l A^2_l\sum_l B_l^2$ in the RHS of Eq.~(\ref{min}), we obtain,  
\be 
\label{j0in}
\label{in3}
 \sum_k\, p_k\,\sum_i \left(\frac{\langle {\vec \sigma_i\cdot {\hat n}_{i0}} \rangle_k}{2}\right)^2\geq \frac{\langle {\cal J}_0\rangle^2}{N}.
\ee
Here, we have substituted (see Eq.~(\ref{gen}) for the definition of ${\cal J}_0$)
\be
\langle {\cal J}_0\rangle^2=\left(\sum_k\, p_k\,\sum_i \frac{\langle{\vec \sigma_i\cdot {\hat n}_{i0}\rangle_k}}{2}\right)^2
\ee 
in a separable state Eq.~(\ref{sp})).

From Eqs.~(\ref{in1}), (\ref{in2}) and (\ref{in3}) we infer that
\be
\left( \Delta{\cal J}_\bot \right)^2\geq\frac{1}{N}\langle {\cal J}_0 \rangle^2\Rightarrow \tilde \xi_2=\frac{{\sqrt N}\Delta {\cal J}_{\bot}}{\vert\langle {\cal J}_0 \rangle\vert }\geq 1
\ee
in a separable $N$ qubit state Eq.~(\ref{sp}). In other words, spin squeezing characterized by $\tilde{\xi}_2\leq 1$ necessarily signifies entanglement. 

\subsection*{\normalsize 5.3 Connection between local invariants and Kitagawa-Ueda spin squeezing in symmetric multiqubit states} 
It is well known that entanglement of a composite quantum system remain invariant, when the subsystems are subjected to local unitary operations~\cite{ushasudha, akrusha}. Any two quantum states are entanglementwise equivalent iff they are related to each other through local unitary transformations. In fact, the non-local properties associated with a quantum state can be represented in terms of a complete set of local invariants~\cite{us2}. While Makhlin \cite{makhlin} had proposed a complete set of 18 local invariants for an arbitrary two-qubit mixed system, it was shown in Ref.~\cite{us2} that the number of invariants reduce to 6, when the two qubit state obeys exchange symmetry. Symmetric  states indeed offer elegant mathematical analysis as the dimension of the Hilbert space  reduces drastically  from $2^N$ to $(N+1)$, when $N$ two-level systems respect exchange symmetry.  It has been shown that all spin squeezed multiatom states are pairwise entangled~\cite{sanders1,ushasudha,ulamku}. Efforts to connect the concept of spin squeezing to the theory of entanglement witnesses~\cite{entwit} have also been carried out in 
Ref.~\cite{korbicz}. Moreover,  \textit{generalized spin squeezing inequalities}, have been proposed~\cite{korbicz} to provide necessary and sufficient conditions for pairwise entanglement and three party entanglement in symmetric $N$-qubit states.   

In fact, any collective phenomena, like spin-squeezing, reflecting pairwise entanglement of symmetric qubits, should be expressible in terms of two qubit local invariants. In fact,  the generalized Kitegawa-Ueda spin squeezing parameter Eq.~(\ref{xitil1}) can be expressed~\cite{ushasudha} in terms of one of the local invariant quantity ${\cal I}$ associated with the symmetric two-qubit reduced density matrix of the $N$ qubit symmetric state.  
We proceed to describe this elegant connection between spin squeezing and pairwise entanglement.    

Among the six local invariants of a symmetric two-qubit state~\cite{ushasudha,us2}, we focus our attention here on one of the invariants,  
\be
{\cal I}=\epsilon_{ijk}\epsilon_{lmn}s_i s_l t_{jm} t_{kn}
\ee
where $\epsilon_{ijk}(\equiv \epsilon_{lmn})$ is the Levi-Civita symbol, $s_i$ are the components of the mean spin vector 
$\vec s$ and $t_{ij}$ are the elements of the two-qubit correlation matrix ${\cal T}$.
 
It has been identified in Ref.~\cite{us2} that ${\cal I}<0$ signifies pairwise entanglement in symmetric multiqubit systems. It is interesting that negative values of the invariant ${\cal I}$ necessarily correspond to spin squeezing characterized by $\tilde{\xi}_1<1$. This connection is brought out explicitly as follows. 

As ${\cal I}$ is invariant under identical local unitary operations, it is convenient to express it after subjecting the quantum state under consideration to an identical local rotation which will align the average spin vector $\vec s$ along the z-axis in which situation, we have ${\vec s}\equiv (0,\,0,\,s_0)$, $s_0=\vert\vec{s}\vert$. So, the invariant ${\cal I}$ gets simplified as follows: 
\be
\label{imp'}
{\cal I}=\epsilon_{3jk}\epsilon_{3mn}s_0^2  t_{jm} t_{kn}=2\,s_0^2(t_{11}t_{22}-t_{12}^2)=2\,s_0^2 \,\mbox{det}\,{\cal T}_\bot
\ee
where ${T}_\bot$ denotes the $2\times 2$ block of the correlation matrix. Exploiting the freedom of local rotations in the x--y plane, leaving the average spin ${\vec s}\equiv (0,\,0,\,s_0)$ unaffected, it is possible to diagonalize ${T}_{\bot}$. We therefore obtain, 
\be
\label{lm}
{\cal T}_{\bot}\equiv \ba{ll} t^{(+)}_{\bot} & 0 \\ 0 & t^{(-)}_{\bot} \ea, \ \ t^{(\pm)}_{\bot}=\frac{1}{2}\left[(t_{11}+t_{22})\pm\sqrt{(t_{11}-t^2_{22})+4 t_{12}^2} \right], 
\ee
where $t^{(\pm)}_{\bot}$ denote the maximum (minimum) eigenvalues of $T_{\bot}$. 
So, the local invariant  ${\cal I}$ assumes a simple form, 
\be
\label{imp}
{\cal I}=2\,s_0^2\,t^{(+)}_{\bot}\,t^{(-)}_{\bot}.
\ee 
Evidently, ${\cal I}$ assumes negative values when $t_{\bot}^{(-)}<0$ --  provided that $t_{\bot}^{(+)}$ is positive. Indeed $t_{\bot}^{(+)}\geq 0$, which may be seen by  recalling that the positivity of any arbitrary two-qubit density matrix imposes the bound~\cite{rh}, $-1\leq t_{11},\, t_{22},\,t_{33}\leq 1$ on the diagonal elements of the correlation matrix $T$ and also the result~\cite{us2}, ${\rm Tr}[T]=1$ for all symmetric states -- from which it is clear that only one of the diagonal elements of the two qubit correlation matrix $T$ can be negative. In other words, 
${\cal I}<0 \Longleftrightarrow t^{(-)}_{\bot}<0$. 

We now turn our attention on the Kitagawa-Ueda spin squeezing parameter $\tilde\xi_1$, associated with a symmetric $N$-qubit state (see Eqs.~(\ref{xitil1}), (\ref{nmin})):  
\begin{eqnarray}
\label{nf}
\tilde\xi_1^2&=&1+(N-1)(n_\bot^T\,T\,n_\bot)_{\mbox{min}}\nonumber \\
&=&1+(N-1)\, t_\bot^{(-)},
\end{eqnarray}
where $t_\bot^{(-)}$ is the least eigenvalue of  $T_{\bot}$ (see Eq.~(\ref{lm})).  
Substituting Eqs.~(\ref{nf}) in Eq.~(\ref{imp}), we get
\be
\label{imp1}
{\cal I}=\frac{2\,s_0^2\,t^{(+)}_{\bot} }{(N-1)}(\tilde\xi_1^2-1).
\ee 
From Eq.~(\ref{imp1}) it is evident that $${\cal I}<0\Longleftrightarrow \tilde{\xi}_1<1$$ 
i.e., a symmetric multiqubit state is spin squeezed  if and only if the local invariant ${\cal I}<0$ -- which establishes the link between  spin squeezing and pairwise entanglement.

\section*{\normalsize 6 Conclusion}
There has been a rapid growth in the field of quantum correlated multiatomic systems, where spin squeezing of an ensemble of a few ions to around $10^7$ cold atoms is being routinely achieved experimentally~\cite{cesium}. Spin squeezing and quantum entanglement do exhibit intrinsic connections with each other -- eventhough they appear to arise from different physical origins. We have discussed in detail the desirable features exhibited by any  measure of spin squeezing and their connection with quantum entanglement in multiqubit systems.

\renewcommand{\refname}{
\end{document}